\documentclass[%
 aip,
 sd,%
 amsmath,amssymb,
 reprint,%
]{revtex4-1}

\usepackage{graphicx}% Include figure files
\usepackage{dcolumn}% Align table columns on decimal point
\usepackage{bm}% bold math
\usepackage{epstopdf}
\begin{document}

\preprint{AIP/123-QED}

\title[]{Optical evaluation of the wave filtering properties of graded undulated lattices}% Force line breaks with \\
%\thanks{Footnote to title of article.}

\author{G. Trainiti}\email{gtrainiti@gatech.edu.}
\affiliation{ 
Georgia Institute of Technology, Daniel Guggenheim School of Aerospace Engineering, Atlanta, 30332, USA
}
\author{M. Ruzzene}%\altaffiliation[Also at ]{Georgia Institute of Technology, George W. Woodruff School of Mechanical Engineering, Atlanta, 30332, USA.}
%\altaffiliation[Also at ]{Georgia Institute of Technology, George W. Woodruff School of Mechanical Engineering, Atlanta, 30332, USA.}%
\affiliation{ 
Georgia Institute of Technology, Daniel Guggenheim School of Aerospace Engineering, Atlanta, 30332, USA
}
\affiliation{Georgia Institute of Technology, George W. Woodruff School of Mechanical Engineering, Atlanta, 30332, USA%\\This line break forced with \textbackslash\textbackslash
}%
% \homepage{http://www.Second.institution.edu/~Charlie.Author.}
%\affiliation{%
%Second institution and/or address%\\This line break forced% with \\
%}%

\date{\today}% It is always \today, today,

\begin{abstract}
We investigate and experimentally demonstrate the elastic wave filtering properties of graded undulated lattices. Square reticulates composed of curved beams are characterized by graded mechanical properties which result from the spatial modulation of the curvature parameter. Among such properties, the progressive formation of frequency bandgaps leads to strong wave attenuation over a broad frequency range. The experimental investigation of wave transmission, and the detection of full wavefields effectively illustrate this behavior. Transmission measurements are conducted using a scanning Laser vibrometer, while a dedicated digital image correlation procedure is implemented to capture in-plane wave motion at selected frequencies. The presented results illustrate the broadband attenuation characteristics resulting from spatial grading of the lattice curvature, whose in-depth investigation is enabled by the presented experimental procedures.
%
%We numerically verify and experimentally validate the elastic wave filtering capabilities of graded structural lattices. This work focuses on undulated structures, obtained by imposing a non-zero value of curvature in square reticulates. Smooth gradation of the local measure of curvature guarantees effective mechanical insulation well outside the band gaps ranges in conventional periodic lattice structures. The design of undulated graded structures relies on a numerical implementation of Bloch analysis in periodic structures. During the experimental testing of 3D printed lattices, we first employ scanning laser Doppler vibrometry to compare the measured transmissibility of the structure to the theoretical one, then a full-field measurement of the in-plane motion of the lattice is obtained through an optical approach based on high-speed camera and digital image correlation.

%
%Valid PACS numbers may be entered using the \verb+\pacs{#1}+ command.
\end{abstract}

%Use showkeys class option if keyword
\maketitle

%\begin{quotation}
%The ``lead paragraph'' is encapsulated with the \LaTeX\ 
%\verb+quotation+ environment and is formatted as a single paragraph before the first section heading. 
%(The \verb+quotation+ environment reverts to its usual meaning after the first sectioning command.) 
%Note that numbered references are allowed in the lead paragraph.
%%
%The lead paragraph will only be found in an article being prepared for the journal \textit{Chaos}.
%\end{quotation}
\section{Introduction}
Structural lattices, obtained tessellating 2D and 3D space with slender beam elements, can be regarded as a special class of mechanical metamaterials, whose properties have been copiously investigated in recent years~\cite{brillouin1953wave,hussein2014dynamics,christensen_kadic_kraft_wegener_2015}. Demand of lightweight and high strength materials, driven by automotive and aerospace industries, has motivated the effort of populating previously forbidden regions of the material properties charts~\cite{Fleck2495}. In lattice materials stiffness, strength and fracture response have been shown to depend upon geometry and nodal connectivity, with behaviors ranging from bending-dominated in foams to stretching-dominated in highly connected cellular solids such as octet-truss lattices~\cite{Fleck2495,Montemayor2015,DESHPANDE20011747}. Recent advancements in fabrication capabilities have further spurred on the interested in fully exploiting architected materials' potential, exploring nanometer-scale lattices~\cite{Bauer2016} as well as hierarchical geometries~\cite{Mousanezhad2015}. Structural metamaterials are perhaps even more appealing for their dynamic properties. Frequency-dependent forbidden elastic wave propagation and strongly directional behavior have been investigated in several lattice topologies and geometries~\cite{phani2006wave,SPADONI2009435,Wang2014}, with possible applications in noise reduction, vibration control and stress wave mitigation~\cite{hussein2014dynamics}. Recently, enhanced functionality in lattices has been explored through nonlinearity to achieve amplitude-dependent response~\cite{Ganesh2017apl2}, tunable directivity with piezoelectric patches and shunted negative capacitance circuits~\cite{Celli2015}, and large deformations effects~\cite{Pal2016}. Another rather unexplored research direction in structural metamaterials is given by graded configurations, in which the periodic repetition of the same unit cell is replaced by a smooth grading of material properties or geometrical features. In this regard, gradient-index phononic crystals (GRIN PCs) can be designed to provide a refractive index profile able to focus elastic energy, realizing acoustic lenses~\cite{PhysRevB.79.094302}, with recent promising extensions to piezoelectric energy harvesting~\cite{Tol2016,Tol2017}. In structural lattices, grading the curvature of the beam elements has been numerically explored in undulated configurations~\cite{Trainiti2016}.
Although it was theoretically shown that structural metamaterials provide a disparate landscape of wave propagation properties, experimental validation is scarcely documented, with most of tests  performed to obtain transmissibility measures through a small number of sensors~\cite{Warmuth2017,DAlessandro2016}. When full wavefield measurement is required, 3D scanning laser Doppler vibrometry is commonly used to measure the wave velocity field in  predefined lattice locations~\cite{CELLI2014,Ganesh2017apl1}. This approach presents a number of shortcomings, mainly related to the cost of the experimental apparatus and the challenge of focusing three laser beams onto the same measurement location, especially for small lattice beams. We recently proposed a different non-contact optical technique to achieve high-spatially resolved wavefield measurements, which is based on optical measurement of the motion through high speed cameras and digital image correlation~\cite{schaeffer2017optical}.
We previously observed that non-periodic, graded configurations display enhanced filtering properties compared to the ones of their periodic counterparts~\cite{Trainiti2016}. In this work, we discuss the experimental investigation of such filtering properties, obtained with and improved version of our optical, digital image correlation-based technique. 
\section{Methods}
\begin{figure}
    \includegraphics[width=0.7\textwidth]{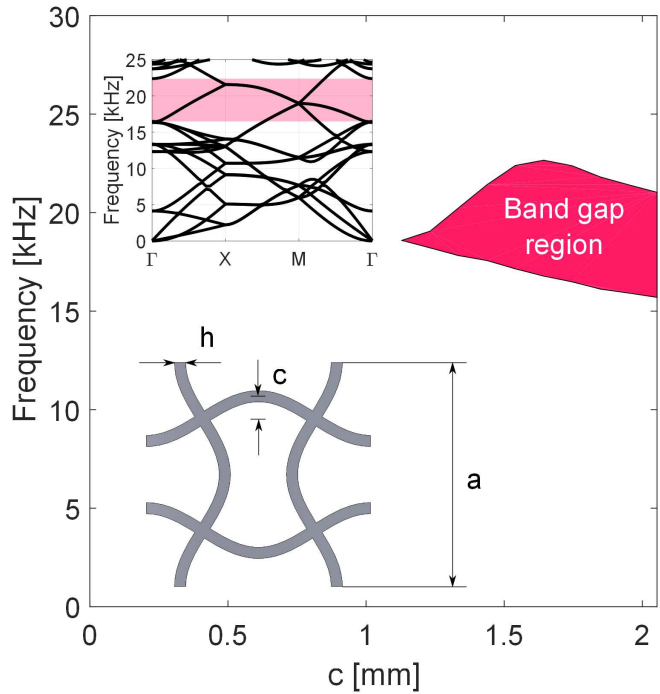}
      \caption{Band gap map of periodic undulated lattices for increasing $c$ with $h=1$ mm. The inset on the bottom left shows the unit cell and its dimensions. The inset on the top left shows the band diagram corresponding to $c=1.75$ mm and how the band gap (in red) is identified. The dispersion branches within the band gap are associated to out-of-plane modes which are not meaningful in an in-plane study. }
      \label{fig:BandGapMap_dim_edited_bis}
      %\centering
\end{figure}

The design of the graded structure relies on a numerical study of equivalent periodic undulated lattices, achieved by analyzing the band gaps locations of infinite periodic structures through a numerical implementation of Bloch's analysis~\citep{aaberg1997usage}. For the experimental validation, we first use scanning laser Doppler vibrometry (SLDV) to obtain transmissibility maps. Such maps illustrate the cumulative effect of the increasing undulation along the structure, informing the design of graded structures. Furthermore, a deeper understanding of the wave propagation phenomenon and the role of undulation gradation in filtering elastic waves is achieved with a second approach, based on the measurement of the in-plane wavefield in both lattices with a high speed camera. In this second experiment, we record the motion of the structure and assume that, at each recorded frame, such motion induces a small perturbation in the pixel intensities of the recorder digital images. Then, we correlate the digital images to indirectly infer conclusions on the structure's motion.
\begin{figure}
    \includegraphics[width=0.7\textwidth]{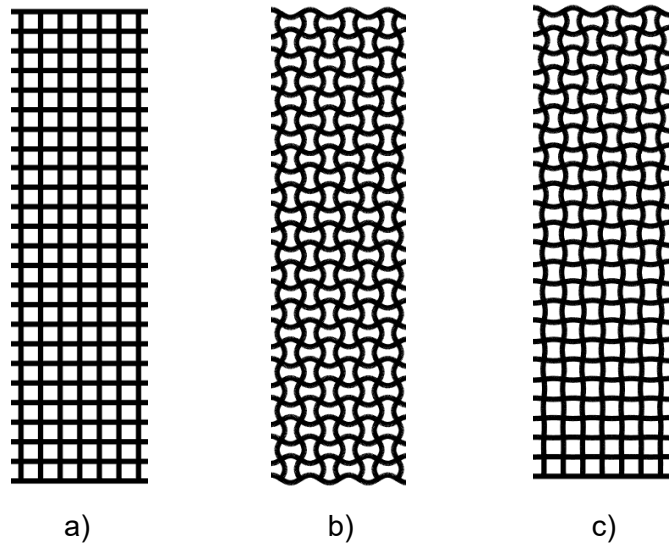}
      \caption{Different lattice configurations: straight (a), periodic undulated (b), graded undulated (c).}
      \label{fig:lattices_schematic}
      %\centering
\end{figure}
Fig.\ref{fig:lattices_schematic} illustrates the structural lattices considered in this study. Undulated structures are obtained by imposing an initial curvature to the linking elements of a square, periodic straight reticulate. The undulation produces a periodic lattice if the imposed curvature varies periodically throughout the structure, while a graded, non-periodic structure is the result of a modulated curvature profile. Among all the possible undulated configurations, we focus our attention on the one inspired by the instability-induced pattern transformation in porous soft materials, due to its interesting wave propagation properties\cite{PhysRevB.77.052105,Trainiti2016}. We start considering an infinite undulated periodic lattice, whose unit cell is represented in Fig.~\ref{fig:BandGapMap_dim_edited_bis}. The geometry of the unit cell is defined by $a$, which is twice the distance between two neighboring lattice interactions, $h$ the thickness of the lattice's beams and $c$ the undulation amplitude.
%We introduce the dimensionless undulation parameter $\zeta$ and the dimensionless thickness $\gamma$, defined as:
%\begin{equation}
%\zeta=\frac{c}{a}, \quad \quad \quad \gamma=\frac{h}{a}
%\end{equation}
In this work, we assume $a=20.5$ mm and we target lattices with slender beams, therefore we consider $h=1$ mm. Also, intending to 3D print and test the lattice structures, we consider the lattice's material to be ABSplus-P430 with tensile modulus $E = 2.2$ GPa and density $\rho=1040$ $\mathrm{kg/m^3}$. We perform a dispersion analysis of the structure by implementing a FE-based Bloch analysis~\cite{aaberg1997usage} and modeling the unit cell with Abaqus C3D6 6-node linear triangular prism elements~\cite{AbaqusDocumentation}. By sweeping the undulation amplitude $c$ from zero (straight lattice) to $c_{max}\approx2$ mm, we construct the band gap map of in-plane wave propagation in Fig.~\ref{fig:BandGapMap_dim_edited_bis}, obtained by identifying the width of the main band gap for each value of the considered $c$. The main band gap appears for $c>c_{cr}$, with $c_{cr}$ a certain critical value which in general depends on the slenderness of the beam, thus on the ratio $h/a$. The widest band gap within the considered range of $c$ is obtained approximately for $c\approx1.6$ mm and it is about $5.5$ kHz wide. In contrast, due the shape of the band gap region in Fig.~\ref{fig:BandGapMap_dim_edited_bis}, band gaps range from $15.7$ kHz to $22.7$ kHz, thus covering a wider  $7$ kHz frequency range. Therefore, one can speculate that a non-periodic structure with smooth graded undulation would benefit from the cumulative contributions of the local value of undulation, leading to an augmented elastic wave filtering capability compared to its periodic counterpart. Experimental validation of graded lattices performance is discussed in the second part of the present work, where tests are performed on two different lattices, a straight and a graded one.

We fabricate the lattices with a total of $N_{U}=12$ unit cells each using a Stratasys Fortus 250mc 3D printer. The value $N_{U}$ is chosen in order to guarantee a smooth linear grading of the undulation in the range $c\in[0,2]$ mm, given the manufacturing constraints in terms of available printing area ($254 \times 254$ mm). During the fabrication process, the lattices were lying flat on the 3D printer building surface, so to minimize induced anisotropy due to uneven material deposition, which would bias the lattice's in-plane dynamics. For the same reason, the highest degree of fill-to-void ratio was imposed.
\section{Results and discussion}
\subsection{Transmissibility maps}
\begin{figure}
    \includegraphics[width=0.7\textwidth]{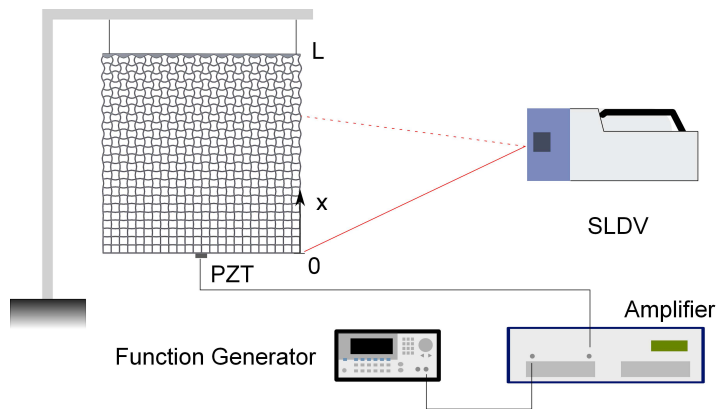}
      \caption{Experimental setup for the measurement of transmissibility maps $T(x,f)$.}
      \label{fig:Exp_schematic_T}
      %\centering
\end{figure}

\begin{figure}
    \includegraphics[width=0.7\textwidth]{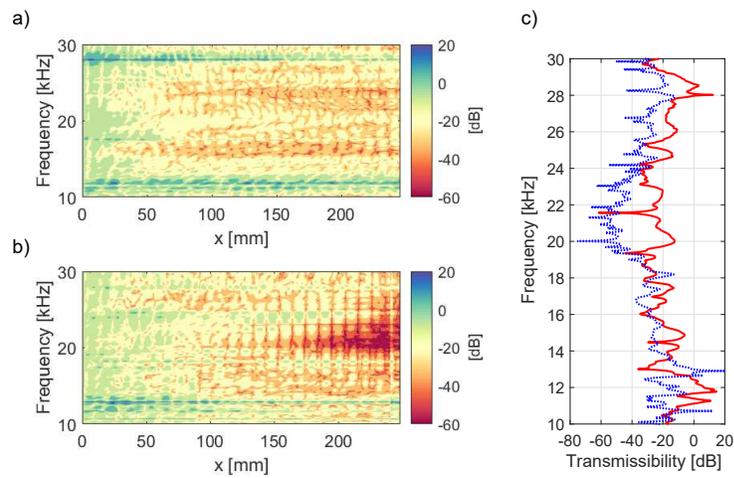}
      \caption{Transmissibility maps $T(x,f)$ for the straight (a) and graded (b) lattices. The graded configuration guarantees a wide transmissibility drop from $20$ to $27.5$ kHz. The difference in transmissibility between the straight (solid red line) and graded (dotted blue line) lattices is shown in detail in (c) for $x=23.5$ mm. }
      \label{fig:T_maps}
      %\centering
\end{figure}

We use the experimental apparatus shown in Fig.~\ref{fig:Exp_schematic_T} to measure the effect of the undulation gradation in the wave propagation attenuation. 
The lattice is hanging from a frame hold by thin cables to approximate a free boundary conditions. The excitation is provided by a piezoelectric disk glued to lattices' edges. The piezoelectric disk generates a broadband signal, then the scanning head of the SLDV measures the transient response of the structure at different locations $x$ along its edge, where $x$ is a reference frame whose origin corresponds to the edge of the graded lattice with zero undulation. Due to the undulated edge's surface, a retroreflective tape is applied to improve the laser's signal quality. The response is recorded in the form of velocity, with sampling rate of $256$ kHz for $8$ ms at each of 400 equally spaced locations from $x=0$ to $x=L=246$  mm. We define the transmissibility map $T(x,f)$, function of the frequency $f$ and the space variable $x$, as the ratio:
\begin{equation}
T(x,f) = 20 \log_{10}\bigg[\frac{s(x,f)}{s(0,f)}\bigg]
\end{equation}
where $s(x,f)$ is the Fourier transform of the signal recorded at the location $x$, and $s(0,f)$ is the Fourier transform of the reference signal recorded at $x=0$. A comparison between the transmissibility maps of the straight and graded lattices is shown in Fig~\ref{fig:T_maps}. The maps show that the graded lattice achieves a dramatic drop of transmissibility (between $\sim40$dB and $\sim60$dB) in the range between $20$ kHz and $27.5$ kHz at $x=L$, thus providing a $7.5$ kHz wide wave attenuation range. Moreover, such drop in transmissibility is particularly visible at $20$ kHz for $x>150$  mm, frequency at which the graded lattice is most effective in filtering elastic waves. In comparing the experimental and theoretical results, we remark that analysis of the band gap map predicts wave attenuation in a $7$ kHz wide range of frequencies, which is in excellent agreement with experimental validation. On the other hand, such range is shifted upwards in frequency, which might be explained partly by the uncertainty in material properties of the 3D printed material, especially the effective elastic modulus.
%\begin{figure}
%    \includegraphics[width=0.5\textwidth]{Figures/T_maps_s_and_g_jet_rev_64_edited_text.eps}
%      \caption{Transmissibility maps for the straight (top) and graded (bottom) lattices.}
%      \label{fig:T_maps}
%      %\centering
%\end{figure}

%\section{Experimental validation}
\begin{figure}
    \includegraphics[width=0.7\textwidth]{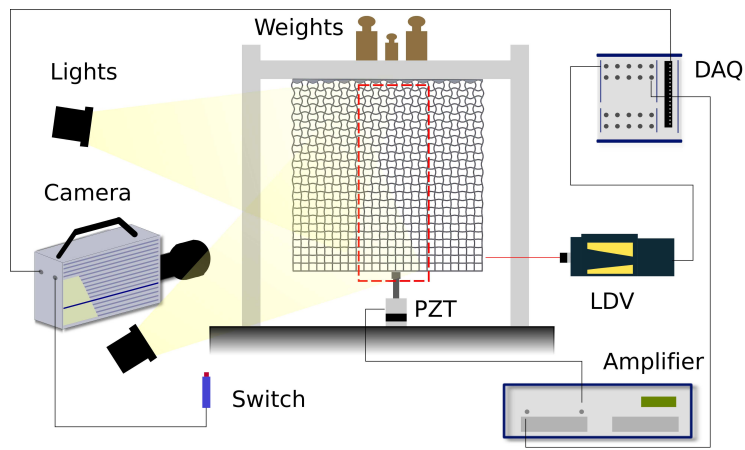}
      \caption{Experimental setup for in-plane wavefield optical measurement with high-speed camera.}
      \label{fig:Exp_schematic_ok}
      %\centering
\end{figure}
\subsection{DIC-based wavefield measurement}
\begin{figure}
    \includegraphics[width=0.70\textwidth]{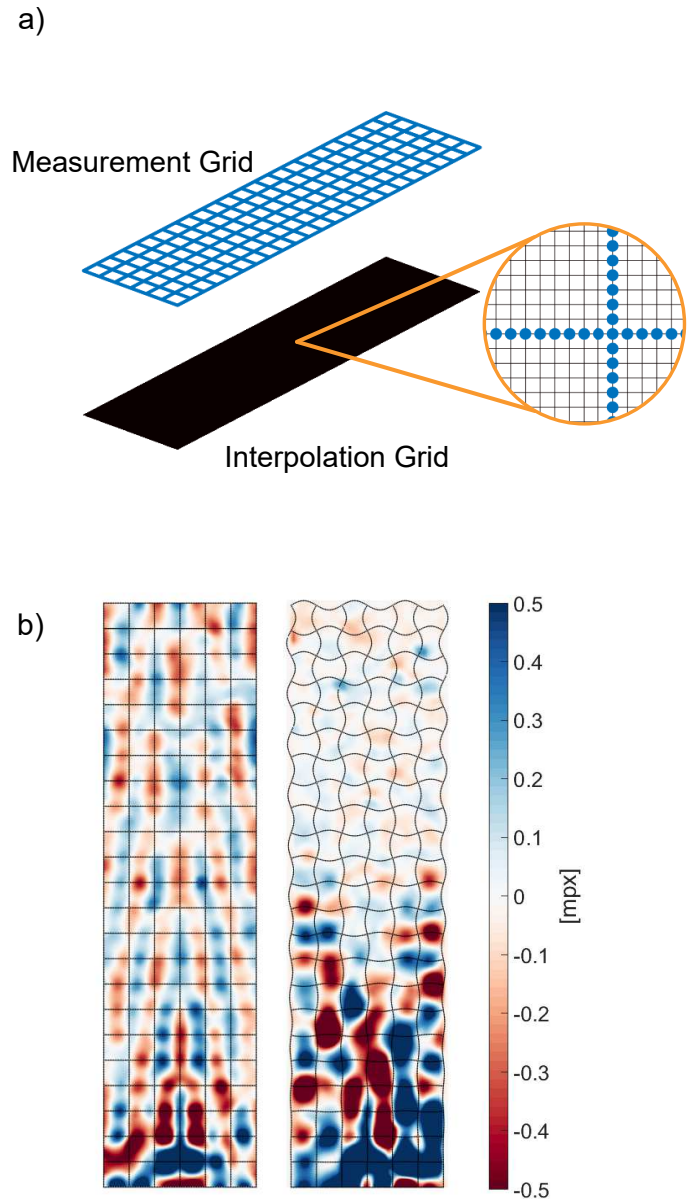}
      \caption{Measurement grid and interpolation grid (a). Interpolated wavefield in the straight and graded lattices show how waves are filtered by the increasing local curvature (b). Black lines represent the measurement locations. Elastic waves are attenuated by the undulation gradation.}
      \label{fig:Interp_map_edited_plus_wavefields}
      %\centering
\end{figure}

Based on the information given by the transmissibility map for the graded lattice, we design a second experiment to measure the in-plane wavefield. We target the response of the system at $20$ kHz, frequency at which the drop of transmissibility is the largest. The experimental setup, shown in Fig.~\ref{fig:Exp_schematic_ok}, employs a single high-speed camera (Photron Fastcam SA1.1) to record the motion of the structure. Adequate light is provided by two high intensity lights (Lowel Pro-light). Elastic waves are excited by actuating an ultrasonic piezo-transducer (APC 90-4060) tuned to resonate at $f=20$ kHz. A $2.5$ kg preload is applied to the structure to improve its coupling with the actuator. For each measurement, a manual switch triggers the recording, then the high-speed camera sends a signal to a data acquisition unit (DAQ, NI USB-6356). Upon receiving the signal from the high-speed camera, the DAQ generates the excitation signal, which is then sent to the amplifier and finally to the actuator. Concurrently, the DAQ records the voltage output from a laser Doppler vibrometer (LDV, Polytec PDV 100), which monitors the structure's response at one point on the lattice's side. In choosing the size of the measurement area, we have to consider the high-speed camera's limitations in reading and storing the information while recording. Higher sampling rates force us to reduce the frame size, thus the number of the recorded pixels within the same image, while choosing larger frame sizes implies reducing the frame rate. This inverse relationship between frame rate and frame size would prevent us from measuring large enough wavefields with a sufficient temporal resolution. Nevertheless, such limitations are overcome by effectively increasing both spatial and temporal sampling, as discussed in our previous works~\cite{schaeffer2017optical,darnton2017optical}. The measurement domain is divided into $23$ tiles, each corresponding to an array of $2\times14$ lattice intersections. For each tile, the experiment is performed separately, then the data is stitched together to retrieve the full wavefield. We choose to track $20$ evenly spaced points between each intersection, which sums up to roughly 8600 measurement points for the total considered measurement area. The experiments are performed at sampling rate $f_s=75$ kHz. An effective sampling rate $f_{s,eff} = 2 f_s = 150$ kHz is realized by properly interleaving two different sets of measurements, which differ by a certain delay $t_d=1/(2 f_s)$ between the beginning of the camera measurement and onset of the excitation. The delay time $t_d$ is imposed by conveniently programming the DAQ. For each set of measurement, we consider 5 averages which help improving the signal-to-noise ratio by reducing the uncorrelated noise. We target the behavior of the system to a narrowband excitation, thus we excite the system with a $11-$cycle tone-burst at $20$ kHz. In order to obtain a sufficiently large excitation signal, the ultrasonic piezo transducer is coupled to a resonator to amplify its response. The piezo-resonator assembly is tuned to have its first resonant frequency at $20$ kHz by properly selecting the resonator's length and conveniently preloading the assembly. The raw data, in the form of pixel intensity variations over the different frames, is collected and pre-processed by correlating the frame set, averaging and interleaving the two measurement sets~\cite{schaeffer2017optical}. The pre-processed data is then post-processed by filtering the data in the frequency domain around the excitation frequency to improve the signal-to-noise ratio. Owing to the highly discontinuous nature of lattice structure, the results visualization is considerably improved by interpolating the data onto a rectangular $121\times461$ grid of points, as shown in Fig.~\ref{fig:Interp_map_edited_plus_wavefields}(a). Finally, a moving average filter is applied to the interpolated data to produce the plots shown in Fig.~\ref{fig:Interp_map_edited_plus_wavefields}(b), which shows the wavefields in both straight and graded configurations, together with the original measurement points at different time instants. We can successfully track the wavefront from the excitation location to the opposite side of the structure. We also remark on the strong directional nature of the wavefront, as expected for straight lattices. The wavefield of the graded lattice, on the other hand, starts differing quite remarkably from the one in the straight lattice already few intersections away from the excitation location, as the effect of the undulation gets strong enough. This effect becomes remarkable halfway through the lattice, preventing the energy carried by the elastic waves from propagating any further, effectively confining it to the first half of the structure.
\section{Conclusion}
In conclusion, we experimentally validated the in-plane filtering properties of undulated lattices, showing that single-phase structural metamaterials can be conveniently designed in graded configurations by slowly varying the curvature of the lattice elements along one direction. We showed that the design of the graded lattice is informed by the dispersion analysis of infinite periodic undulated lattice, thorough inspection of the band gap map representing the relation between the band gap width and the beam element's curvature. Then, we compute transmissibility maps by measuring the velocity field at one edge of a graded and an equivalent straight lattice with an SLDV system, identifying a frequency range with transmission attenuation in the graded configuration. Finally, we employ a high speed camera and digital image correlation to measure the full wavefield for a narrowband excitation with frequency spectrum falling within the large attenuation frequency range in both lattices, tracking how in-plane elastic waves are attenuated in the graded one only due to the gradual change in geometry. Future research directions include extending the full wavefield measurement herein discussed for the study of the directional properties of graded lattices, with applications in superior energy guiding, energy focusing and energy harvesting.
%\begin{figure}
%    \includegraphics[width=0.5\textwidth]{Figures/DIC_S_cut.eps}
%      \caption{Interpolated wavefield in the straight lattice. Black lines represent the measurement locations. Elastic waves excited at one end of the structure propagate along the structure.}
%      \label{fig:DIC_S_cut}
%      %\centering
%\end{figure}
%
%\begin{figure}
%    \includegraphics[width=0.5\textwidth]{Figures/DIC_G_cut.eps}
%      \caption{Interpolated wavefield in the graded lattice. Black lines represent the measurement locations. Elastic waves are attenuated by the undulation gradation.}
%      \label{fig:DIC_G_cut}
%      %\centering
%\end{figure}

%\begin{figure}
%    \includegraphics[width=0.5\textwidth]{Figures/DIC_SandG_cut.eps}
%      \caption{Interpolated wavefield in the straight (a) and graded (b) lattices. Black lines represent the measurement locations. Elastic waves are attenuated by the undulation gradation.}
%      \label{fig:DIC_SandG_cut}
%      %\centering
%\end{figure}

%\section{Graded Lattice design}
%\section{Graded lattice design and fabrication}
%The lattices we test in this study are made of ABSplus-P430 and were 3D printed by using a Stratasys Fortus 250mc. The lattice's design is based on the material specifications provided by the manufacturer, \emph{i.e.} tensile modulus $E = 2.2$ GPa and density $\rho=1040$ $kg/m^3$. A smooth grading of the undulation requires having a sufficient number $N_{U}$ of unit cells in the equivalent periodic straight lattice. Given the manufacturing constraints in terms of available printing area ($25.4 \times 25.4 \times 30.5 \,\,\,cm$) and the target frequency for capturing the wave filtering properties of the graded structure, we assumed $N_{U}=12$, spacing between two neighboring intersections $a=2.05\,\,\,cm$ and thickness of the lattice's beams $h = 1.03\,\,\,cm$. It also follows that the maximum undulation amplitude in the graded lattice is given by $c_0=\zeta a= 0.2\,\,\,cm$. During the fabrication process, the lattices were laying flat on the 3D printer building surface, so to minimize induced anisotropy due to uneven material deposition, which would bias the lattice's in-plane dynamics. For the same reason, the highest degree of fill-to-void ratio was imposed. 

\begin{acknowledgments}
The work is supported by the National Science Foundation - CMMI/ENG Division, through grant 1719728.
\end{acknowledgments}

\nocite{*}
\bibliography{references_}% Produces the bibliography via BibTeX.

\begin{thebibliography}{28}%
\makeatletter
\providecommand \@ifxundefined [1]{%
 \@ifx{#1\undefined}
}%
\providecommand \@ifnum [1]{%
 \ifnum #1\expandafter \@firstoftwo
 \else \expandafter \@secondoftwo
 \fi
}%
\providecommand \@ifx [1]{%
 \ifx #1\expandafter \@firstoftwo
 \else \expandafter \@secondoftwo
 \fi
}%
\providecommand \natexlab [1]{#1}%
\providecommand \enquote  [1]{``#1''}%
\providecommand \bibnamefont  [1]{#1}%
\providecommand \bibfnamefont [1]{#1}%
\providecommand \citenamefont [1]{#1}%
\providecommand \href@noop [0]{\@secondoftwo}%
\providecommand \href [0]{\begingroup \@sanitize@url \@href}%
\providecommand \@href[1]{\@@startlink{#1}\@@href}%
\providecommand \@@href[1]{\endgroup#1\@@endlink}%
\providecommand \@sanitize@url [0]{\catcode `\\12\catcode `\$12\catcode
  `\&12\catcode `\#12\catcode `\^12\catcode `\_12\catcode `\%12\relax}%
\providecommand \@@startlink[1]{}%
\providecommand \@@endlink[0]{}%
\providecommand \url  [0]{\begingroup\@sanitize@url \@url }%
\providecommand \@url [1]{\endgroup\@href {#1}{\urlprefix }}%
\providecommand \urlprefix  [0]{URL }%
\providecommand \Eprint [0]{\href }%
\providecommand \doibase [0]{http://dx.doi.org/}%
\providecommand \selectlanguage [0]{\@gobble}%
\providecommand \bibinfo  [0]{\@secondoftwo}%
\providecommand \bibfield  [0]{\@secondoftwo}%
\providecommand \translation [1]{[#1]}%
\providecommand \BibitemOpen [0]{}%
\providecommand \bibitemStop [0]{}%
\providecommand \bibitemNoStop [0]{.\EOS\space}%
\providecommand \EOS [0]{\spacefactor3000\relax}%
\providecommand \BibitemShut  [1]{\csname bibitem#1\endcsname}%
\let\auto@bib@innerbib\@empty
%</preamble>
\bibitem [{\citenamefont {Brillouin}(1953)}]{brillouin1953wave}%
  \BibitemOpen
  \bibfield  {author} {\bibinfo {author} {\bibfnamefont {L.}~\bibnamefont
  {Brillouin}},\ }\href@noop {} {\emph {\bibinfo {title} {Wave propagation in
  periodic structures: electric filters and crystal lattices}}},\ Dover books
  and science\ (\bibinfo  {publisher} {Dover Publications},\ \bibinfo {year}
  {1953})\BibitemShut {NoStop}%
\bibitem [{\citenamefont {Hussein}, \citenamefont {Leamy},\ and\ \citenamefont
  {Ruzzene}(2014)}]{hussein2014dynamics}%
  \BibitemOpen
  \bibfield  {author} {\bibinfo {author} {\bibfnamefont {M.~I.}\ \bibnamefont
  {Hussein}}, \bibinfo {author} {\bibfnamefont {M.~J.}\ \bibnamefont {Leamy}},
  \ and\ \bibinfo {author} {\bibfnamefont {M.}~\bibnamefont {Ruzzene}},\
  }\bibfield  {title} {\enquote {\bibinfo {title} {Dynamics of phononic
  materials and structures: Historical origins, recent progress, and future
  outlook},}\ }\href@noop {} {\bibfield  {journal} {\bibinfo  {journal} {Appl.
  Mech. Rev.}\ }\textbf {\bibinfo {volume} {66}},\ \bibinfo {pages} {040802}
  (\bibinfo {year} {2014})}\BibitemShut {NoStop}%
\bibitem [{\citenamefont {Christensen}\ \emph {et~al.}(2015)\citenamefont
  {Christensen}, \citenamefont {Kadic}, \citenamefont {Kraft},\ and\
  \citenamefont {Wegener}}]{christensen_kadic_kraft_wegener_2015}%
  \BibitemOpen
  \bibfield  {author} {\bibinfo {author} {\bibfnamefont {J.}~\bibnamefont
  {Christensen}}, \bibinfo {author} {\bibfnamefont {M.}~\bibnamefont {Kadic}},
  \bibinfo {author} {\bibfnamefont {O.}~\bibnamefont {Kraft}}, \ and\ \bibinfo
  {author} {\bibfnamefont {M.}~\bibnamefont {Wegener}},\ }\bibfield  {title}
  {\enquote {\bibinfo {title} {Vibrant times for mechanical metamaterials},}\
  }\href {\doibase 10.1557/mrc.2015.51} {\bibfield  {journal} {\bibinfo
  {journal} {MRS Communications}\ }\textbf {\bibinfo {volume} {5}},\ \bibinfo
  {pages} {453–462} (\bibinfo {year} {2015})}\BibitemShut {NoStop}%
\bibitem [{\citenamefont {Fleck}, \citenamefont {Deshpande},\ and\
  \citenamefont {Ashby}(2010)}]{Fleck2495}%
  \BibitemOpen
  \bibfield  {author} {\bibinfo {author} {\bibfnamefont {N.~A.}\ \bibnamefont
  {Fleck}}, \bibinfo {author} {\bibfnamefont {V.~S.}\ \bibnamefont
  {Deshpande}}, \ and\ \bibinfo {author} {\bibfnamefont {M.~F.}\ \bibnamefont
  {Ashby}},\ }\bibfield  {title} {\enquote {\bibinfo {title}
  {Micro-architectured materials: past, present and future},}\ }\href@noop {}
  {\bibfield  {journal} {\bibinfo  {journal} {Proceedings of the Royal Society
  of London A: Mathematical, Physical and Engineering Sciences}\ }\textbf
  {\bibinfo {volume} {466}},\ \bibinfo {pages} {2495--2516} (\bibinfo {year}
  {2010})}\BibitemShut {NoStop}%
\bibitem [{\citenamefont {Montemayor~LC}(2015)}]{Montemayor2015}%
  \BibitemOpen
  \bibfield  {author} {\bibinfo {author} {\bibfnamefont {G.~J.}\ \bibnamefont
  {Montemayor~LC}},\ }\bibfield  {title} {\enquote {\bibinfo {title}
  {Mechanical response of hollow metallic nanolattices: Combining structural
  and material size effects},}\ }\href@noop {} {\bibfield  {journal} {\bibinfo
  {journal} {Journal of Applied Mechanics}\ }\textbf {\bibinfo {volume} {82}},\
  \bibinfo {pages} {071012--071012--10} (\bibinfo {year} {2015})}\BibitemShut
  {NoStop}%
\bibitem [{\citenamefont {Deshpande}, \citenamefont {Fleck},\ and\
  \citenamefont {Ashby}(2001)}]{DESHPANDE20011747}%
  \BibitemOpen
  \bibfield  {author} {\bibinfo {author} {\bibfnamefont {V.}~\bibnamefont
  {Deshpande}}, \bibinfo {author} {\bibfnamefont {N.}~\bibnamefont {Fleck}}, \
  and\ \bibinfo {author} {\bibfnamefont {M.}~\bibnamefont {Ashby}},\ }\bibfield
   {title} {\enquote {\bibinfo {title} {Effective properties of the octet-truss
  lattice material},}\ }\href@noop {} {\bibfield  {journal} {\bibinfo
  {journal} {Journal of the Mechanics and Physics of Solids}\ }\textbf
  {\bibinfo {volume} {49}},\ \bibinfo {pages} {1747 -- 1769} (\bibinfo {year}
  {2001})}\BibitemShut {NoStop}%
\bibitem [{\citenamefont {Bauer}\ \emph {et~al.}(2016)\citenamefont {Bauer},
  \citenamefont {Schroer}, \citenamefont {Schwaiger},\ and\ \citenamefont
  {Kraft}}]{Bauer2016}%
  \BibitemOpen
  \bibfield  {author} {\bibinfo {author} {\bibfnamefont {J.}~\bibnamefont
  {Bauer}}, \bibinfo {author} {\bibfnamefont {A.}~\bibnamefont {Schroer}},
  \bibinfo {author} {\bibfnamefont {R.}~\bibnamefont {Schwaiger}}, \ and\
  \bibinfo {author} {\bibfnamefont {O.}~\bibnamefont {Kraft}},\ }\bibfield
  {title} {\enquote {\bibinfo {title} {Approaching theoretical strength in
  glassy carbon nanolattices},}\ }\href@noop {} {\bibfield  {journal} {\bibinfo
   {journal} {Nature Materials}\ ,\ \bibinfo {pages} {438–443}} (\bibinfo
  {year} {2016})}\BibitemShut {NoStop}%
\bibitem [{\citenamefont {ans Babaee S ans Ebrahimi H ans Ghosh R ans Hamouda
  AS ans Bertoldi K~ans Ashkan~Vaziri}(2015)}]{Mousanezhad2015}%
  \BibitemOpen
  \bibfield  {author} {\bibinfo {author} {\bibfnamefont {M.~D.}\ \bibnamefont
  {ans Babaee S ans Ebrahimi H ans Ghosh R ans Hamouda AS ans Bertoldi K~ans
  Ashkan~Vaziri}},\ }\bibfield  {title} {\enquote {\bibinfo {title}
  {Hierarchical honeycomb auxetic metamaterials},}\ }\href@noop {} {\bibfield
  {journal} {\bibinfo  {journal} {Scientific Reports}\ }\textbf {\bibinfo
  {volume} {5}},\ \bibinfo {pages} {18306} (\bibinfo {year}
  {2015})}\BibitemShut {NoStop}%
\bibitem [{\citenamefont {Phani}, \citenamefont {Woodhouse},\ and\
  \citenamefont {Fleck}(2006)}]{phani2006wave}%
  \BibitemOpen
  \bibfield  {author} {\bibinfo {author} {\bibfnamefont {A.~S.}\ \bibnamefont
  {Phani}}, \bibinfo {author} {\bibfnamefont {J.}~\bibnamefont {Woodhouse}}, \
  and\ \bibinfo {author} {\bibfnamefont {N.}~\bibnamefont {Fleck}},\ }\bibfield
   {title} {\enquote {\bibinfo {title} {Wave propagation in two-dimensional
  periodic lattices},}\ }\href@noop {} {\bibfield  {journal} {\bibinfo
  {journal} {J. Acoust. Soc. Am.}\ }\textbf {\bibinfo {volume} {119}},\
  \bibinfo {pages} {1995--2005} (\bibinfo {year} {2006})}\BibitemShut {NoStop}%
\bibitem [{\citenamefont {Spadoni}\ \emph {et~al.}(2009)\citenamefont
  {Spadoni}, \citenamefont {Ruzzene}, \citenamefont {Gonella},\ and\
  \citenamefont {Scarpa}}]{SPADONI2009435}%
  \BibitemOpen
  \bibfield  {author} {\bibinfo {author} {\bibfnamefont {A.}~\bibnamefont
  {Spadoni}}, \bibinfo {author} {\bibfnamefont {M.}~\bibnamefont {Ruzzene}},
  \bibinfo {author} {\bibfnamefont {S.}~\bibnamefont {Gonella}}, \ and\
  \bibinfo {author} {\bibfnamefont {F.}~\bibnamefont {Scarpa}},\ }\bibfield
  {title} {\enquote {\bibinfo {title} {Phononic properties of hexagonal chiral
  lattices},}\ }\href@noop {} {\bibfield  {journal} {\bibinfo  {journal} {Wave
  Motion}\ }\textbf {\bibinfo {volume} {46}},\ \bibinfo {pages} {435 -- 450}
  (\bibinfo {year} {2009})}\BibitemShut {NoStop}%
\bibitem [{\citenamefont {Wang}, \citenamefont {Wang},\ and\ \citenamefont
  {Zhang}(2014)}]{Wang2014}%
  \BibitemOpen
  \bibfield  {author} {\bibinfo {author} {\bibfnamefont {Y.-F.}\ \bibnamefont
  {Wang}}, \bibinfo {author} {\bibfnamefont {Y.-S.}\ \bibnamefont {Wang}}, \
  and\ \bibinfo {author} {\bibfnamefont {C.}~\bibnamefont {Zhang}},\ }\bibfield
   {title} {\enquote {\bibinfo {title} {Bandgaps and directional properties of
  two-dimensional square beam-like zigzag lattices},}\ }\href@noop {}
  {\bibfield  {journal} {\bibinfo  {journal} {AIP Advances}\ }\textbf {\bibinfo
  {volume} {4}},\ \bibinfo {pages} {124403} (\bibinfo {year}
  {2014})}\BibitemShut {NoStop}%
\bibitem [{\citenamefont {Ganesh}\ and\ \citenamefont
  {Gonella}(2017{\natexlab{a}})}]{Ganesh2017apl2}%
  \BibitemOpen
  \bibfield  {author} {\bibinfo {author} {\bibfnamefont {R.}~\bibnamefont
  {Ganesh}}\ and\ \bibinfo {author} {\bibfnamefont {S.}~\bibnamefont
  {Gonella}},\ }\bibfield  {title} {\enquote {\bibinfo {title} {Experimental
  evidence of directivity-enhancing mechanisms in nonlinear lattices},}\
  }\href@noop {} {\bibfield  {journal} {\bibinfo  {journal} {Applied Physics
  Letters}\ }\textbf {\bibinfo {volume} {110}},\ \bibinfo {pages} {084101}
  (\bibinfo {year} {2017}{\natexlab{a}})}\BibitemShut {NoStop}%
\bibitem [{\citenamefont {Celli}\ and\ \citenamefont
  {Gonella}(2015)}]{Celli2015}%
  \BibitemOpen
  \bibfield  {author} {\bibinfo {author} {\bibfnamefont {P.}~\bibnamefont
  {Celli}}\ and\ \bibinfo {author} {\bibfnamefont {S.}~\bibnamefont
  {Gonella}},\ }\bibfield  {title} {\enquote {\bibinfo {title} {Tunable
  directivity in metamaterials with reconfigurable cell symmetry},}\
  }\href@noop {} {\bibfield  {journal} {\bibinfo  {journal} {Applied Physics
  Letters}\ }\textbf {\bibinfo {volume} {106}},\ \bibinfo {pages} {091905}
  (\bibinfo {year} {2015})}\BibitemShut {NoStop}%
\bibitem [{\citenamefont {Raj Kumar~Pal}\ and\ \citenamefont
  {Ruzzene}(2016)}]{Pal2016}%
  \BibitemOpen
  \bibfield  {author} {\bibinfo {author} {\bibfnamefont {J.~R.}\ \bibnamefont
  {Raj Kumar~Pal}}\ and\ \bibinfo {author} {\bibfnamefont {M.}~\bibnamefont
  {Ruzzene}},\ }\bibfield  {title} {\enquote {\bibinfo {title} {Tunable
  directivity in metamaterials with reconfigurable cell symmetry},}\
  }\href@noop {} {\bibfield  {journal} {\bibinfo  {journal} {Smart Materials
  and Structures}\ }\textbf {\bibinfo {volume} {25}},\ \bibinfo {pages}
  {054010} (\bibinfo {year} {2016})}\BibitemShut {NoStop}%
\bibitem [{\citenamefont {Lin}\ \emph {et~al.}(2009)\citenamefont {Lin},
  \citenamefont {Huang}, \citenamefont {Sun},\ and\ \citenamefont
  {Wu}}]{PhysRevB.79.094302}%
  \BibitemOpen
  \bibfield  {author} {\bibinfo {author} {\bibfnamefont {S.-C.~S.}\
  \bibnamefont {Lin}}, \bibinfo {author} {\bibfnamefont {T.~J.}\ \bibnamefont
  {Huang}}, \bibinfo {author} {\bibfnamefont {J.-H.}\ \bibnamefont {Sun}}, \
  and\ \bibinfo {author} {\bibfnamefont {T.-T.}\ \bibnamefont {Wu}},\
  }\bibfield  {title} {\enquote {\bibinfo {title} {Gradient-index phononic
  crystals},}\ }\href {\doibase 10.1103/PhysRevB.79.094302} {\bibfield
  {journal} {\bibinfo  {journal} {Phys. Rev. B}\ }\textbf {\bibinfo {volume}
  {79}},\ \bibinfo {pages} {094302} (\bibinfo {year} {2009})}\BibitemShut
  {NoStop}%
\bibitem [{\citenamefont {Tol}, \citenamefont {Degertekin},\ and\ \citenamefont
  {Erturk}(2016)}]{Tol2016}%
  \BibitemOpen
  \bibfield  {author} {\bibinfo {author} {\bibfnamefont {S.}~\bibnamefont
  {Tol}}, \bibinfo {author} {\bibfnamefont {F.~L.}\ \bibnamefont {Degertekin}},
  \ and\ \bibinfo {author} {\bibfnamefont {A.}~\bibnamefont {Erturk}},\
  }\bibfield  {title} {\enquote {\bibinfo {title} {Gradient-index phononic
  crystal lens-based enhancement of elastic wave energy harvesting},}\
  }\href@noop {} {\bibfield  {journal} {\bibinfo  {journal} {Applied Physics
  Letters}\ }\textbf {\bibinfo {volume} {109}},\ \bibinfo {pages} {063902}
  (\bibinfo {year} {2016})}\BibitemShut {NoStop}%
\bibitem [{\citenamefont {Tol}, \citenamefont {Degertekin},\ and\ \citenamefont
  {Erturk}(2017)}]{Tol2017}%
  \BibitemOpen
  \bibfield  {author} {\bibinfo {author} {\bibfnamefont {S.}~\bibnamefont
  {Tol}}, \bibinfo {author} {\bibfnamefont {F.~L.}\ \bibnamefont {Degertekin}},
  \ and\ \bibinfo {author} {\bibfnamefont {A.}~\bibnamefont {Erturk}},\
  }\bibfield  {title} {\enquote {\bibinfo {title} {Phononic crystal luneburg
  lens for omnidirectional elastic wave focusing and energy harvesting},}\
  }\href {\doibase 10.1063/1.4991684} {\bibfield  {journal} {\bibinfo
  {journal} {Applied Physics Letters}\ }\textbf {\bibinfo {volume} {111}},\
  \bibinfo {pages} {013503} (\bibinfo {year} {2017})}\BibitemShut {NoStop}%
\bibitem [{\citenamefont {Trainiti}, \citenamefont {Rimoli},\ and\
  \citenamefont {Ruzzene}(2016)}]{Trainiti2016}%
  \BibitemOpen
  \bibfield  {author} {\bibinfo {author} {\bibfnamefont {G.}~\bibnamefont
  {Trainiti}}, \bibinfo {author} {\bibfnamefont {J.}~\bibnamefont {Rimoli}}, \
  and\ \bibinfo {author} {\bibfnamefont {M.}~\bibnamefont {Ruzzene}},\
  }\bibfield  {title} {\enquote {\bibinfo {title} {Wave propagation in
  undulated structural lattices},}\ }\href@noop {} {\bibfield  {journal}
  {\bibinfo  {journal} {International Journal of Solids and Structures}\
  }\textbf {\bibinfo {volume} {97-98}},\ \bibinfo {pages} {431--444} (\bibinfo
  {year} {2016})}\BibitemShut {NoStop}%
\bibitem [{\citenamefont {Warmuth}, \citenamefont {Wormser},\ and\
  \citenamefont {Körner}(2017)}]{Warmuth2017}%
  \BibitemOpen
  \bibfield  {author} {\bibinfo {author} {\bibfnamefont {F.}~\bibnamefont
  {Warmuth}}, \bibinfo {author} {\bibfnamefont {M.}~\bibnamefont {Wormser}}, \
  and\ \bibinfo {author} {\bibfnamefont {C.}~\bibnamefont {Körner}},\
  }\bibfield  {title} {\enquote {\bibinfo {title} {Single phase 3d phononic
  band gap material},}\ }\href@noop {} {\bibfield  {journal} {\bibinfo
  {journal} {Scientific Reports}\ }\textbf {\bibinfo {volume} {7}},\ \bibinfo
  {pages} {3843} (\bibinfo {year} {2017})}\BibitemShut {NoStop}%
\bibitem [{\citenamefont {D'Alessandro}\ \emph {et~al.}(2016)\citenamefont
  {D'Alessandro}, \citenamefont {Belloni}, \citenamefont {Ardito},
  \citenamefont {Corigliano},\ and\ \citenamefont {Braghin}}]{DAlessandro2016}%
  \BibitemOpen
  \bibfield  {author} {\bibinfo {author} {\bibfnamefont {L.}~\bibnamefont
  {D'Alessandro}}, \bibinfo {author} {\bibfnamefont {E.}~\bibnamefont
  {Belloni}}, \bibinfo {author} {\bibfnamefont {R.}~\bibnamefont {Ardito}},
  \bibinfo {author} {\bibfnamefont {A.}~\bibnamefont {Corigliano}}, \ and\
  \bibinfo {author} {\bibfnamefont {F.}~\bibnamefont {Braghin}},\ }\bibfield
  {title} {\enquote {\bibinfo {title} {Modeling and experimental verification
  of an ultra-wide bandgap in 3d phononic crystal},}\ }\href@noop {} {\bibfield
   {journal} {\bibinfo  {journal} {Applied Physics Letters}\ }\textbf {\bibinfo
  {volume} {109}},\ \bibinfo {pages} {221907} (\bibinfo {year}
  {2016})}\BibitemShut {NoStop}%
\bibitem [{\citenamefont {Celli}\ and\ \citenamefont
  {Gonella}(2014)}]{CELLI2014}%
  \BibitemOpen
  \bibfield  {author} {\bibinfo {author} {\bibfnamefont {P.}~\bibnamefont
  {Celli}}\ and\ \bibinfo {author} {\bibfnamefont {S.}~\bibnamefont
  {Gonella}},\ }\bibfield  {title} {\enquote {\bibinfo {title} {Laser-enabled
  experimental wavefield reconstruction in two-dimensional phononic
  crystals},}\ }\href {\doibase http://dx.doi.org/10.1016/j.jsv.2013.09.001}
  {\bibfield  {journal} {\bibinfo  {journal} {Journal of Sound and Vibration}\
  }\textbf {\bibinfo {volume} {333}},\ \bibinfo {pages} {114 -- 123} (\bibinfo
  {year} {2014})}\BibitemShut {NoStop}%
\bibitem [{\citenamefont {Ganesh}\ and\ \citenamefont
  {Gonella}(2017{\natexlab{b}})}]{Ganesh2017apl1}%
  \BibitemOpen
  \bibfield  {author} {\bibinfo {author} {\bibfnamefont {R.}~\bibnamefont
  {Ganesh}}\ and\ \bibinfo {author} {\bibfnamefont {S.}~\bibnamefont
  {Gonella}},\ }\bibfield  {title} {\enquote {\bibinfo {title} {Experimental
  evidence of directivity-enhancing mechanisms in nonlinear lattices},}\
  }\href@noop {} {\bibfield  {journal} {\bibinfo  {journal} {Applied Physics
  Letters}\ }\textbf {\bibinfo {volume} {110}},\ \bibinfo {pages} {084101}
  (\bibinfo {year} {2017}{\natexlab{b}})}\BibitemShut {NoStop}%
\bibitem [{\citenamefont {Schaeffer}, \citenamefont {Trainiti},\ and\
  \citenamefont {Ruzzene}(2017)}]{schaeffer2017optical}%
  \BibitemOpen
  \bibfield  {author} {\bibinfo {author} {\bibfnamefont {M.}~\bibnamefont
  {Schaeffer}}, \bibinfo {author} {\bibfnamefont {G.}~\bibnamefont {Trainiti}},
  \ and\ \bibinfo {author} {\bibfnamefont {M.}~\bibnamefont {Ruzzene}},\
  }\bibfield  {title} {\enquote {\bibinfo {title} {Optical measurement of
  in-plane waves in mechanical metamaterials through digital image
  correlation},}\ }\href@noop {} {\bibfield  {journal} {\bibinfo  {journal}
  {Scientific Reports}\ }\textbf {\bibinfo {volume} {7}},\ \bibinfo {pages}
  {42437} (\bibinfo {year} {2017})}\BibitemShut {NoStop}%
\bibitem [{\citenamefont {{\AA}berg}\ and\ \citenamefont
  {Gudmundson}(1997)}]{aaberg1997usage}%
  \BibitemOpen
  \bibfield  {author} {\bibinfo {author} {\bibfnamefont {M.}~\bibnamefont
  {{\AA}berg}}\ and\ \bibinfo {author} {\bibfnamefont {P.}~\bibnamefont
  {Gudmundson}},\ }\bibfield  {title} {\enquote {\bibinfo {title} {The usage of
  standard finite element codes for computation of dispersion relations in
  materials with periodic microstructure},}\ }\href@noop {} {\bibfield
  {journal} {\bibinfo  {journal} {The Journal of the Acoustical Society of
  America}\ }\textbf {\bibinfo {volume} {102}},\ \bibinfo {pages} {2007--2013}
  (\bibinfo {year} {1997})}\BibitemShut {NoStop}%
\bibitem [{\citenamefont {Bertoldi}\ and\ \citenamefont
  {Boyce}(2008)}]{PhysRevB.77.052105}%
  \BibitemOpen
  \bibfield  {author} {\bibinfo {author} {\bibfnamefont {K.}~\bibnamefont
  {Bertoldi}}\ and\ \bibinfo {author} {\bibfnamefont {M.~C.}\ \bibnamefont
  {Boyce}},\ }\bibfield  {title} {\enquote {\bibinfo {title} {Mechanically
  triggered transformations of phononic band gaps in periodic elastomeric
  structures},}\ }\href@noop {} {\bibfield  {journal} {\bibinfo  {journal}
  {Phys. Rev. B}\ }\textbf {\bibinfo {volume} {77}},\ \bibinfo {pages} {052105}
  (\bibinfo {year} {2008})}\BibitemShut {NoStop}%
\bibitem [{\citenamefont {Hibbitt}(2012)}]{AbaqusDocumentation}%
  \BibitemOpen
  \bibfield  {author} {\bibinfo {author} {\bibfnamefont {S.}~\bibnamefont
  {Hibbitt}, \bibfnamefont {Karlsson}},\ }\href@noop {} {\emph {\bibinfo
  {title} {ABAQUS/Standard Analysis User's Manual}}}\ (\bibinfo  {publisher}
  {Dassault Syst{\`e}mes},\ \bibinfo {year} {2012})\BibitemShut {NoStop}%
\bibitem [{\citenamefont {Darnton}\ and\ \citenamefont
  {Ruzzene}(2017)}]{darnton2017optical}%
  \BibitemOpen
  \bibfield  {author} {\bibinfo {author} {\bibfnamefont {A.~T.}\ \bibnamefont
  {Darnton}}\ and\ \bibinfo {author} {\bibfnamefont {M.}~\bibnamefont
  {Ruzzene}},\ }\bibfield  {title} {\enquote {\bibinfo {title} {Optical
  measurement of guided waves},}\ }\href@noop {} {\bibfield  {journal}
  {\bibinfo  {journal} {The Journal of the Acoustical Society of America}\
  }\textbf {\bibinfo {volume} {141}},\ \bibinfo {pages} {EL465--EL469}
  (\bibinfo {year} {2017})}\BibitemShut {NoStop}%
\bibitem [{\citenamefont {Trainiti}, \citenamefont {Rimoli},\ and\
  \citenamefont {Ruzzene}(2015)}]{Trainiti2015}%
  \BibitemOpen
  \bibfield  {author} {\bibinfo {author} {\bibfnamefont {G.}~\bibnamefont
  {Trainiti}}, \bibinfo {author} {\bibfnamefont {J.}~\bibnamefont {Rimoli}}, \
  and\ \bibinfo {author} {\bibfnamefont {M.}~\bibnamefont {Ruzzene}},\
  }\bibfield  {title} {\enquote {\bibinfo {title} {Wave propagation in
  periodically undulated beams and plates},}\ }\href@noop {} {\bibfield
  {journal} {\bibinfo  {journal} {International Journal of Solids and
  Structures}\ }\textbf {\bibinfo {volume} {75--76}},\ \bibinfo {pages} {260 --
  276} (\bibinfo {year} {2015})}\BibitemShut {NoStop}%
\end{thebibliography}%

\end{document}